\documentclass[aps,prl,showpacs,twocolumn,superscriptaddress]{revtex4}

\usepackage{graphicx}
\usepackage{dcolumn}
\usepackage{bm}

\begin{document}



\title{Coexistence of Self-Organized Criticality and Intermittent Turbulence in the Solar Corona}

\author{Vadim M. Uritsky}
\email{vuritsky@phas.ucalgary.ca}
\affiliation{Complexity Science Group, Department of Physics and Astronomy,
University of Calgary, Calgary, Alberta, Canada T2N 1N4}

\author{Maya Paczuski}%
\affiliation{Complexity Science Group, Department of Physics and Astronomy,
University of Calgary, Calgary, Alberta, Canada T2N 1N4}

\author{Joseph M. Davila}%
\affiliation{NASA Goddard Space Flight Center, Greenbelt, MD 20771, USA}

\author{Shaela I. Jones}%
\affiliation{University of Maryland, College Park, MD 20742, USA}

\date{\today}

\begin{abstract}

An extended data set of extreme ultraviolet images of the solar corona provided by the SOHO spacecraft are analyzed using statistical methods common to studies of self-organized criticality (SOC) and intermittent turbulence (IT). The data exhibits simultaneous hallmarks of both  regimes, namely power law avalanche statistics as well as multiscaling of structure functions for spatial activity. This implies that both SOC and IT may  be manifestations of a single complex
dynamical process entangling avalanches of magnetic energy dissipation with turbulent particle flows.
  
\end{abstract}
\maketitle

SOC and IT represent two paths to dynamical complexity in driven,
extended nonlinear systems. In classical fluid turbulence, scaling is
often associated with a hierarchical structure of eddies extending
over the inertial range~\cite{Kolm41}, while in SOC, avalanches of
localized instabilities organize the system toward a steady state
exhibiting long-range correlations up to the system size~\cite{Bak87}.
Some authors have argued that SOC and IT are distinct, unrelated
phenomena~\cite{Boff99, Ant01, Carb02}, while others have suggested a
fundamental connection~\cite{Bak05, Chen04, Sreev04, Chang99,
Maya05}. In fact, Bak and co-workers have speculated that
turbulence in the limit of high Reynolds number (HRN) may itself be a
SOC phenomenon~\cite{Bak90,Maya93,Bak05}, with dynamical transitions
observed in HRN turbulence being critical avalanches of dissipation.
It has also been shown~\cite{Maya05} that, in contrast to earlier
claims~\cite{Boff99}, SOC and IT cannot be distinguished by analyzing
interoccurrence times between bursts: once a finite observation
threshold (unavoidable in any physical measurement) is introduced,
even the ordinary BTW sandpile exhibits scale free waiting time
statistics~\cite{Maya05} indicative of turbulent systems.
 
We propose that coexistence of SOC and IT may be a generic feature of magnetized astrophysical plasmas. One scenario for this to occur is localized instabilities ~\cite{Klim04,Dahl05,KUP06} switching the plasma
between frozen and unfrozen states. This process resembles stick-slip or depinning transitions 
of SOC models~\cite{bak89,Maya96} and is to some extent analogous to rice pile dynamics~\cite{Fret96}, where kinetic energy of grains releases stored potential energy. Signatures of SOC and IT have also been
obtained in MHD simulations~\cite{Dmit97,Reale05,Galt98} mimicking footpoint motions of coronal magnetic loops. However, the explicit complementarity between SOC and IT in astrophysical observations has not been demonstrated.

In order to clarify this issue, we present direct observational
evidence for coexistence of SOC and IT in the magnetized plasma of the
solar corona. Using a single high resolution data set, we apply two
different analysis methods -- one for analyzing avalanche statistics
of the emission field and the other for analyzing structure functions
of the same field.  The energy, area and lifetime statistics of
avalanches detected in this data set obey robust scaling laws. Unlike
previous studies of flare statistics, we use a spatiotemporal event
detection algorithm compatible with the usual definition of avalanches
in SOC. Next, we show that the same data set exhibits multiscaling and
extended self-similarity (ESS) of higher order structure functions --
a hallmark of IT phenomena.  The observed scaling laws show only weak
dependence on average solar activity and were found both at solar
minimum (min) and maximum (max), indicating that coexistence of SOC
and IT is a generic characteristic of coronal behavior.

Dissipation mechanisms in the corona are activated by changes in the
configuration of its magnetic field. Convection of magnetic fields
leads to radiative transients, plasma jets, and explosive events known
as flares~\cite{Asch00, Berg99,Delab95}. The latter are associated
with spatially concentrated release of magnetic energy accompanied by
localized plasma heating up to temperatures of $10^7$K, and can be
observed by short-wavelength light emission. Flares tend to appear at
irregular times and locations and exhibit broadband energy, size and
lifetime statistics with no obvious characteristics scales.  This
behavior is often interpreted as a signature of SOC~\cite{Charb01,
  Asch02, Parn00, Berg98, Lu91, Hugh03}.  Further, the configuration
of magnetic flux in the corona has been shown to form a scale-free
network with statistical features that can be captured with a
self-organizing network model~\cite{Hugh03}.  Within active regions, the
magnetic field also has an intermittent spatial structure that
reorganizes during large flares~\cite{Abr03}. Our results suggest,
though, that complementary coexistence of SOC and IT is a generic
phenomenon not limited to these large events.

\begin{figure}[htbp]
\includegraphics[width=8.5cm]{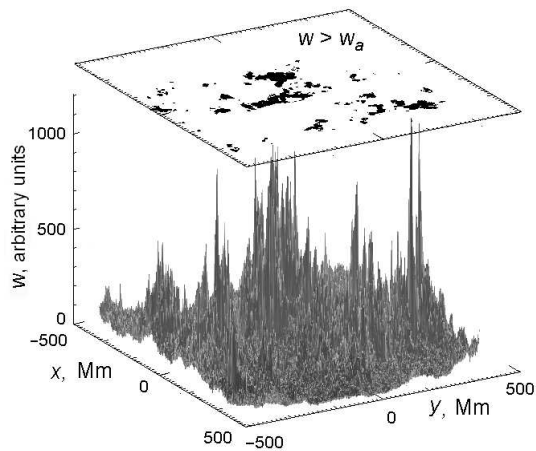}
\vskip -5.0cm
\caption{\label{Fig_1} Two views of the corona's complexity. Upper
  panel: snapshots of high activity coronal regions used to construct
  avalanches ($w_a=350$). Lower panel: a snapshot of the continuous
  brightness field at the same time instant.}
\vskip -0.7cm
\end{figure}

We have studied time series of full-disk digital images of the corona
taken by the extreme ultraviolet imaging telescope (EIT) onboard the
SOHO spacecraft~\cite{Delab95} in the 195 \r{A} wavelength band
corresponding to the Fe XII emission at peak coronal temperatures of
$1.6 \cdot 10^6$ K. The data included two observation periods:
06/29/2001 - 07/28/2001 (3240 images, solar max, average sunspot
number 64.0) and 10/22/2005 - 12/02/2005 (4407 images, solar min,
average sunspot number 16.3) with a typical time resolution of 13.3
min. To reduce optical distortions, we studied only central portion of
the Sun disk with the linear dimensions $1040 \times 1040$ Mm ($256
\times 256$ pixels with 5.6 arcsec resolution). The EIT luminosity
$w(t,\bm{r})$ was analyzed as a function of time $t$ and position
$\bm{r}$ on the image plane. The dynamics of $w(t,\bm{r})$ captures
the redistribution of radiative flux in consecutive EIT frames due to
a variety of coronal features such as
loops and holes, mass ejections, etc.  For the purpose
of this study, we treated $w(t,\bm{r})$ as a local measure of coronal
activity and did not filter the data in an attempt to distinguish
between different types of coronal events. Our analysis was based on
two alternative approaches allowing a study of $w(t,\bm{r})$ both as
an impulsive avalanching process and a continuum turbulent field -- as
shown in Fig.~1.

To identify avalanches, we used a spatiotemporal detection
method~\cite{Urit02, Urit03} that resolves concurrent events. First,
avalanching regions were identified by applying an activity threshold
$w_a$ representing a background EUV flux. Contiguous spatial regions
with $w(\bm{r},t) > w_a$ were treated as pieces of evolving
avalanches. By checking for overlap of common pixels between each pair
of consecutive EIT frames, we identified a set of 3-dimensional
spatiotemporal integration domains $\Lambda_i (i=1,..,N)$
corresponding to each of $N$ individual avalanches. These domains of
contiguous activity in space and time were used to compute the
lifetimes, $T_i=\max(t \in \Lambda_i)-\min (t \in \Lambda_i)$, the
radiative emission flux, $E_i=\int_{\Lambda_i} w(\bm{r},t) \,
d\bm{r}dt$, as well the peak areas, $A_i=\max\limits_{t} (
\int_{\Lambda_i(t)} \, d\bm{r} )$, or maximum number of pixels in a
snapshot of each avalanche. Active regions that split after starting
at a unique source were considered parts of a single avalanche. Active
regions that merged were considered as separate avalanches, with the
common "tail" ascribed to the event that started earlier. Only events
that lasted at least two successive time frames and were not truncated
by the field of view or temporal gaps in observations longer than 40
min were selected for subsequent analysis.  The robustness of the
obtained statistics was verified by repeatedly running the algorithm
with substantially different $w_a$.  Due to large difference in
average emission, $\left\langle w \right\rangle$, during solar min and
max, $w_a$ were defined relative to $\left\langle w \right\rangle$ for
each data set. Depending on $w_a$, between 4,830 (1,680) and 26,900
(5,350) coronal events were detected for solar max (min).

Fig.~2 shows probability distributions for avalanche lifetime, total
emission flux and peak area for both solar max and min.
These statistics can be approximately fitted by the power-law relations
$p(T){\sim}T^{-\tau_T}$, $p(E){\sim}E^{-\tau_E}$ and
$p(A){\sim}A^{-\tau_A}$ with the exponents being almost independent of
$w_a$. The same exponents have also been found using a fluctuating
threshold placed at 3 standard deviations above the average brightness
of each image. The large-scale rollovers are due to the lack of
events whose $T$ approaches the maximum
available time scale $5.2 \cdot 10^5$s given by the ratio between
the latitudinal size of the field of view and the rotation velocity at
the equator. Such events tend to cross the field of view and are therefore underrepresented in our sample.

\begin{figure}[htbp]
\includegraphics*[width=4.2cm]{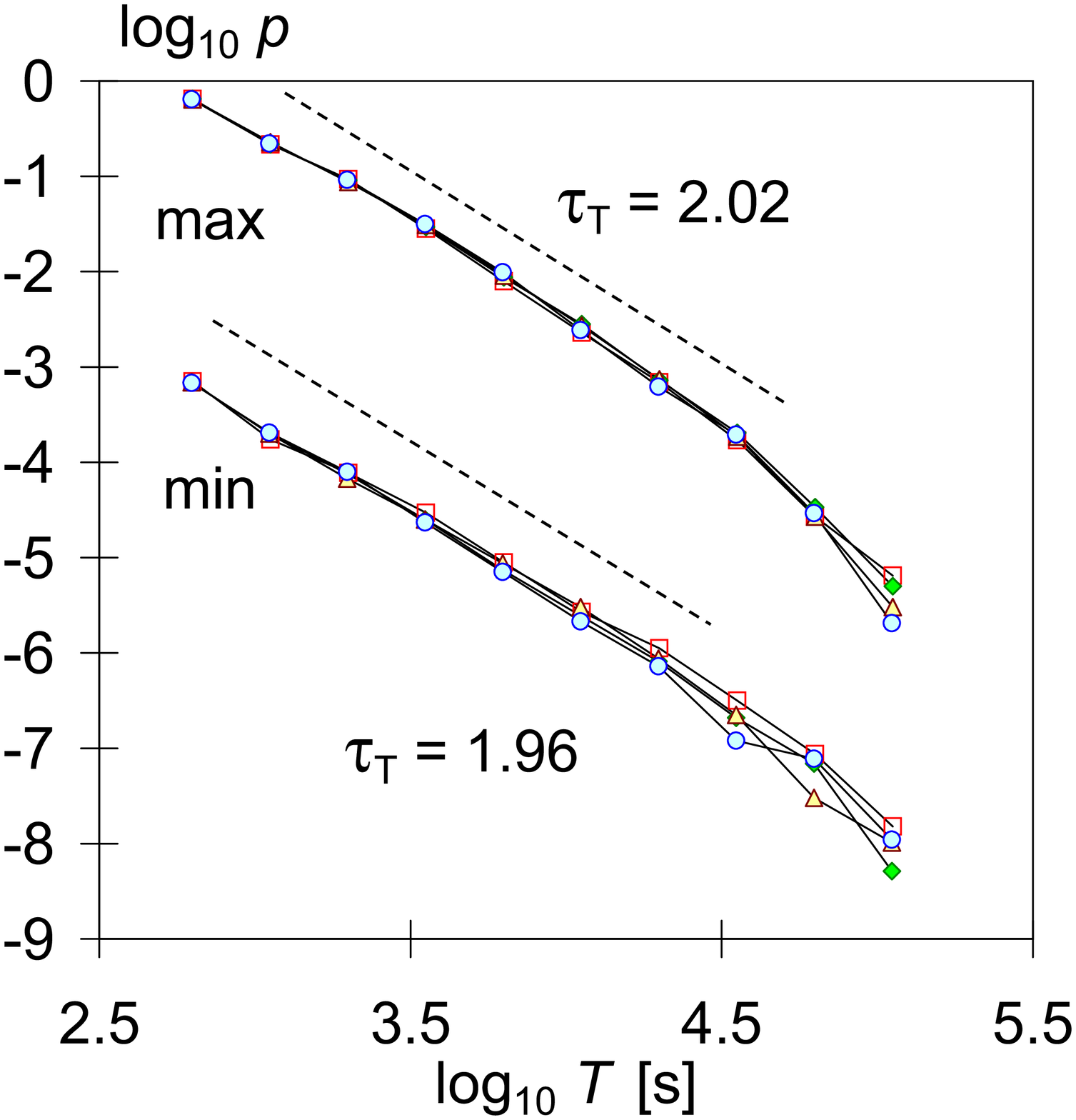}
\includegraphics*[width=4.2cm]{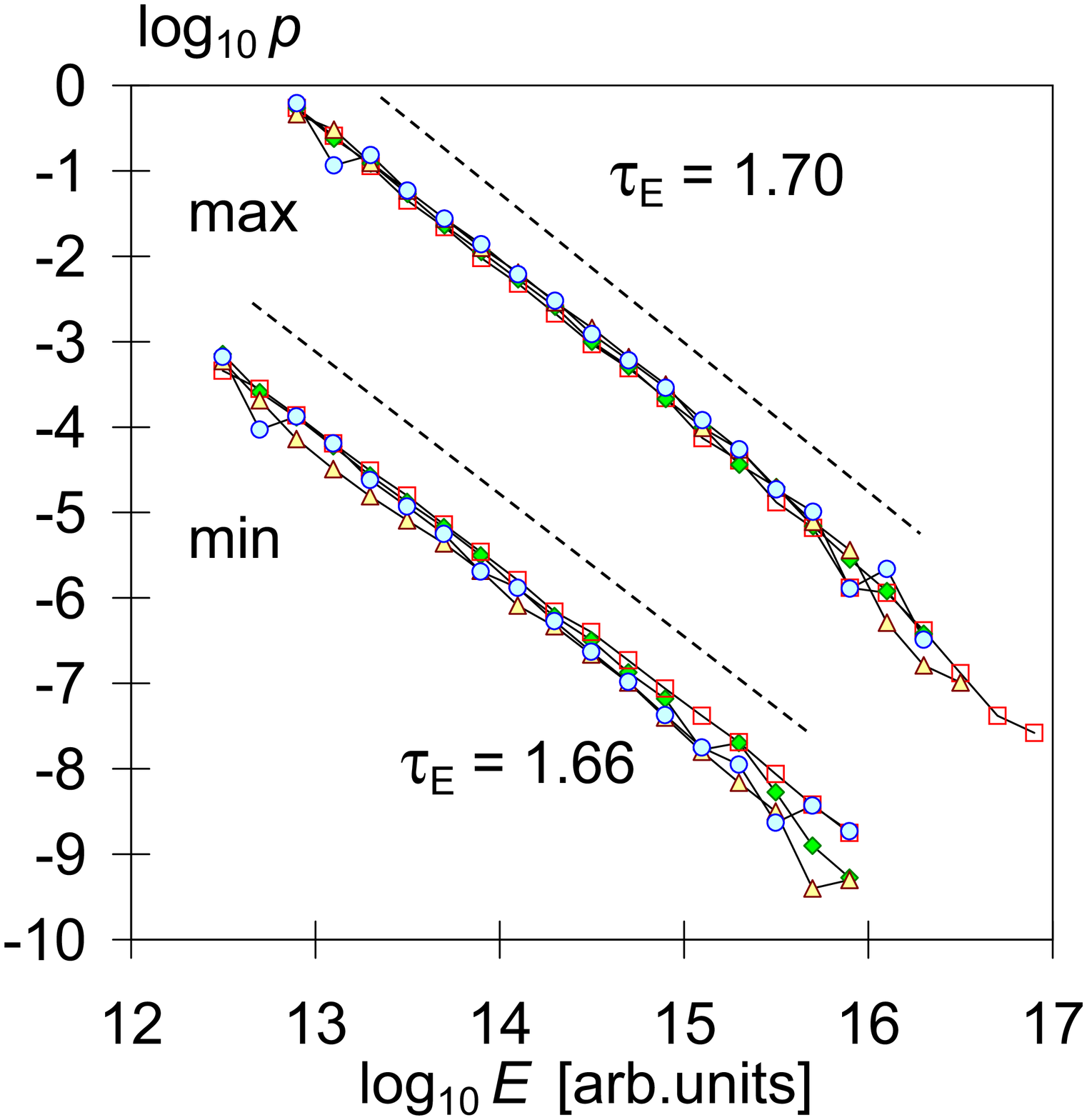}
\vskip -1.2cm
\includegraphics*[width=4.2cm]{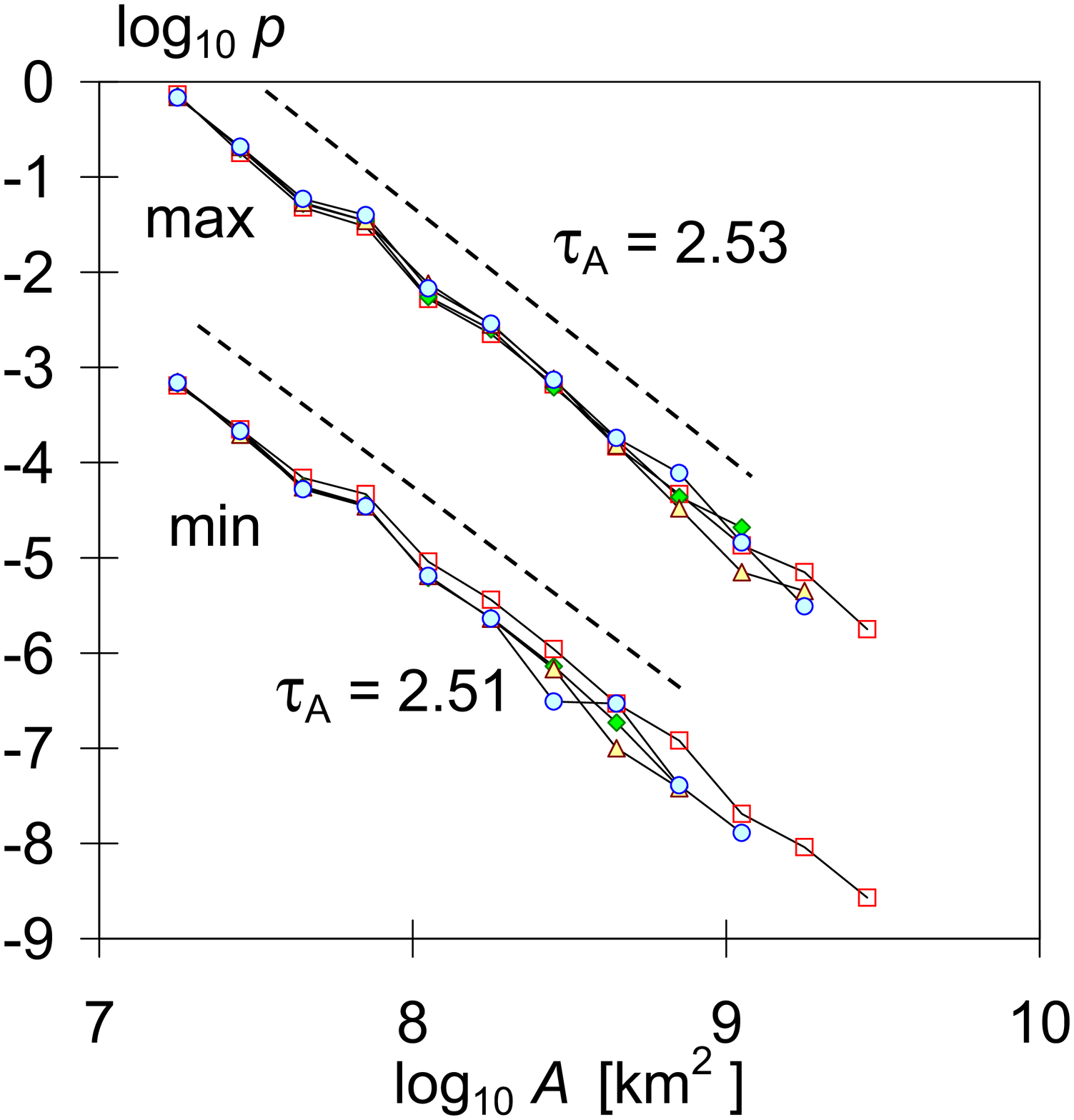}
\vskip -1.3cm

\caption{\label{Fig_2} Probability distributions of avalanche lifetime, emission flux and peak area for solar max and min at four different thresholds $w_a =k\left\langle w \right\rangle$ with $k$ = 0.4 (squares), $k$=0.8 (diamonds), $k$=1.2 (triangles) and $k$=1.6 (circles). The min distributions are shifted downward for comparison.}
\vskip -0.3cm
\end{figure}

The exponents observed at solar max ($\tau_T=2.02\pm0.05$,
$\tau_E=1.70\pm0.04$, $\tau_A=2.53\pm0.09$) and at solar min
($\tau_T=1.96\pm0.06$, $\tau_E=1.66\pm0.03$, $\tau_A=2.51\pm0.14$) are
indistinguishable within uncertainties. This is also true for the
exponents $z$, $D_E$ and $D_A$ defined by the relations $T \sim l^z$,
$E \sim l^{D_E}$ and $A \sim l^{D_A}$, where $l$ is the linear
avalanche scale \footnote{The exponent $D_A$ was estimated by counting
  the number $N_A$ of non-overlapping boxes of size $l$ needed to
  cover the activity at the peak of the avalanche, using $N_A(l)
  \sim l^{-D_A}$. The exponents $z$ and $D_A$ were obtained from
  regression plots $T(A)$ and $E(A)$ assuming $A \sim l^{D_A}$.}.  The
resulting values are $z=1.94\pm0.12$, $D_E=3.16\pm0.21$ and
$D_A=1.50\pm0.10$ for solar max and $z=2.09\pm0.17$, $D_E=3.34\pm0.25$
and $D_A=1.52\pm0.11$ for solar min. All values were obtained by
averaging over four activity thresholds ($w_a=k \left\langle w
\right\rangle$, where $k \in \left\{0.4, 0.8, 1.2, 1.6 \right\} $)
within fixed ranges of scales corresponding to the power-law portions
of the relations. The reported uncertainties are the standard errors
from this averaging or from the regression estimate at individual
thresholds, whichever is larger. Up to these errors, the exponents
satisfy the probability conservation relations $z(\tau_T-1) =
D_E(\tau_E-1) = D_A(\tau_A-1)$.

All these results support the hypothesis that the corona operates in a
SOC state.  It is worth emphasizing that the exponents reported here
have been obtained using spatiotemporal definition of avalanches which
is conceptually much closer to measuring avalanches in numerical simulations than most
of the definitions used in previous works on flare statistics --
except for a few case studies~\cite{Berg98, Berg99} focusing on
specific coronal conditions. The energy distribution exponent $\tau_E$
is smaller than 2 indicating that the coronal heating is
dominated by large events as opposed to Parker's scenario of nanoflare
heating~\cite{Park88}. This conclusion is in agreement with previous
estimates based on spatiotemporal detection of coronal
brightenings~\cite{Berg98, Berg99} and their spatial detection with
subsequent integration of the emission fluxes over fixed time
interval~\cite{Asch00, Asch02}. 
The exponent $D_A$ matches the fractal dimension of
nanoflares reported in~\cite{Asch02}, whereas $\tau_T$ is consistent
with previous analyses of threshold-dependent inter-occurrence times
of x-ray bursts measured over the whole Sun~\cite{Bai06}, as would be
expected if total emission were a sum over individual
avalanches~\cite{Cor99}. Its value as well as $z \approx 2$ indicate
that the corona may operate at a mean-field limit~\cite{Tang88}. We
also note that the avalanche exponents reported are consistent with
the results of forced MHD simulation of turbulent coronal heating~\cite{Dmit97}.


As our next step, we have analyzed coronal activity as a continuum turbulent
field. Such analyses are  normally accomplished by measuring 
structure functions~\cite{Bif05} for a relevant dynamical variable
$\delta z$. Taking $\delta z $ proportional
to differences in the scalar field $w(\bm{r},t)$ ~\cite{Note5}, we define the structure functions of order $q$ to be
\begin{equation}
S_q(l)={\langle| w(\bm{r},t)-w(\bm{r}+\delta \bm{r},t)|^q \rangle}_{\bm{r}}
\label{eq1}
\end{equation}
Here $\delta\bm{r}$ is the spatial displacement, $l\equiv
|\delta\bm{r}|$, and averaging indicated by brackets is performed over
all positions $\bm{r}$. Within the inertial range, $S_q(l) \sim
l^{\zeta(q)}$ with $\zeta(q)$ defined by the turbulent regime under
study. In practice, scaling directly in $S_q(l)$ is often limited in
range. This is exactly the issue we faced when analyzing the coronal
images (Fig.~3).  In such situations, the ESS method~\cite{Benz93} is
often used which allows to extend the observed range of turbulent
scaling making it possible to estimate relative
exponents~\footnote{This is usually done by plotting $S_q(l)$ versus
  $S_3(l)$}. The non-trivial behavior of the moments is approximately
isotropic within uncertainties (see the inset in Fig.~4) and so could
not be eliminated by projecting the data onto irreducible
representations of a lower symmetry group~\cite{Bif05}.

\begin{figure}[htbp]
\includegraphics*[width=8cm]{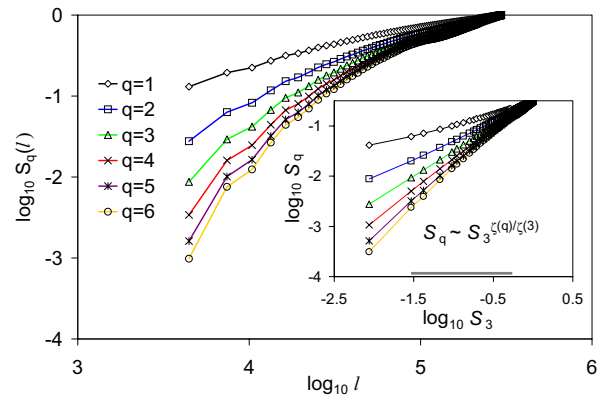}
\vskip -1.0 cm
\caption{\label{Fig_3} Structure functions of coronal activity. Each
  function is normalized by its maximum value. Inset: ESS scaling of
  the structure functions. The horizontal bar shows the range used
  for estimating exponents.}
\end{figure}

\begin{figure}[htbp]
\vskip -0.5cm
\includegraphics*[width=8cm]{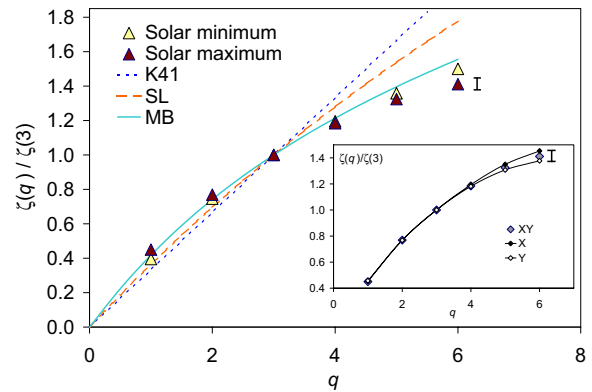}
\vskip -1.0 cm
\caption{\label{Fig_4} ESS exponents as compared to the Kolmogorov
  $q/3$ scaling (K41) and hierarchical models of IT defined in the
  text. Inset: solar maximum exponents obtained for horizontal ($X$),
  vertical ($Y$), and arbitrary ($XY$) orientations of the
  displacement vector $\delta \bm{r}$, as a test for isotropy. Error
  bars in both plots show the discrepancy between horizontal and
  vertical $\zeta(6)$ estimates at solar max.}
\end{figure}

The functions $S_q(l)$ exhibit ESS over almost the entire range
(Fig.~3, inset).  The relative ESS exponent
$\zeta(q)/ \zeta(3)$ (Fig.~4) shows a clear
departure from the Kolmogorov law. Both of these features are typical
of IT systems. To highlight the origin of this intermittency, we have
tested several analytical fits encompassed by a hierarchical
model ~\cite{Polit95, Mull00}:
\begin{equation}
  \zeta(q)=(q/g)(1-x)+C(1-(1-x/C)^{q/g})
\label{eq2}
\end{equation}
The model contains three tuning parameters defined by the relations
$\delta z \sim \ell ^{1/g}$, $t_e \sim \ell ^x$, and $C=3-D$, where
$t_e$ is the energy transfer time at smallest inertial scales $\ell$
and $D$ is the dimension of the dissipative structures. The
She-Leveque (SL) model~\cite{She94} with vortex filaments ($D=1$) is
obtained substituting $g=3$, $x=2/3$, and $C=2$. The
Iroshnikov-Kraichnan MHD model~\cite{Irosh63} assumes $g=4$, $x=1/2$,
and $C=1$ with dissipative structures interpreted as current sheets
and predicts values of the relative exponents which for the $q$ range
considered here are indistinguishable from the SL model. The
combination $g=3$, $x=2/3$, and $C=1$ gives the M\"{u}ller-Biskamp
(MB) model~\cite{Mull00} implying that turbulence occurs in a 3d MHD
system with  hydrodynamic scaling and sheetlike dissipative
structures. As Fig.~4 shows, the latter provides the best overall
description for our data. 

Our main finding is the simultaneous appearance of robust signatures
of both SOC and IT in a single time series of coronal images -- including
significantly different phases of the solar cycle. These observations can be interpreted
using a variety of physical scenarios. One of them assumes that
turbulence in coronal dissipation is passively driven by SOC
avalanches of dissipating currents which modify the
magnetic field and shape its intermittent spatial
pattern~\cite{Abr03}. A more collaborative
interaction between SOC and MHD turbulence is also possible in which SOC avalanches of
reconnecting magnetic loops~\cite{Charb01,Hugh03} generate inward and
outward plasma flows~\cite{Naru06} and/or cascades of MHD shock
waves~\cite{Ryut03} working as sources of turbulence-driven
anomalous resistivity regions which may affect the avalanches \cite{Klim04}

Among the key open problems is the one of the primary physical mechanism of the coronal avalanches. A plausible model has been proposed by Dahlburg et al. \cite{Dahl05} who have shown that secondary instabilities needed to support the avalanche can be triggered by the misalignment between the reconnecting flux tubes. The model is based on sheet-like dissipative structures compatible with the MB turbulent cascade \cite{Mull00}, and predicts rapidly evolving instabilities with the energy dissipation growth time of about $200$ Alfven transit times across the sheets \cite{Dahl05}. These instabilities, reminiscent of toppling events in sandpiles \cite{Bak87}, can propagate throughout extended coronal regions and lead to the multiscale explosive release of the coronal energy accompanied with SOC and IT signatures. Observational verification of this mechanism is a task for future research.


\begin{acknowledgments}
We thank V.~Abramenko, A.~Klimas and A.~Pouquet for valuable discussions.
\end{acknowledgments}

\bibliography{SOHO_short}
\end{document}